%% file: main.tex
\documentclass[conference,9pt]{IEEEtran}
\IEEEoverridecommandlockouts
\usepackage{cite}
\usepackage{amsmath,amssymb,amsfonts}
\usepackage{graphicx}
\usepackage{textcomp}
\usepackage{outlines}
\usepackage{float}
\usepackage{array} 
\usepackage{booktabs} 
\usepackage[table]{xcolor}
\usepackage[justification=centering]{caption}
\usepackage{tikz}
\usepackage{cite}
\usepackage{amsmath,amssymb,amsfonts}
\usepackage{algorithm}
\usepackage{algpseudocode}
\usepackage{graphicx}
\usepackage{subcaption}
\usepackage{textcomp}
\usepackage{xcolor} 
\usepackage{threeparttable} 
\usepackage{multirow}
\usepackage{enumitem}
\usepackage[export]{adjustbox}
\usepackage{tikz}
\usepackage{afterpage} 
\usepackage{color}
\usepackage{pifont}
\usepackage[T1]{fontenc}
\usepackage{array}
\usepackage{booktabs}
\usepackage{float}
\usepackage{stfloats}
\usepackage{hyperref}
\usepackage{listings}
\usepackage{fontawesome5}
\usepackage{notoccite}
\usepackage{arydshln}
\usepackage{vcell}
\usepackage{titlesec}
\usepackage{makecell}
\usepackage{colortbl} 
\usepackage[table]{xcolor}  
\usepackage[moderate, tracking=normal, title=normal, margins=tight, bibliography=normal, lists=normal, floats=tight, paragraphs=normal]{savetrees}
\usepackage{setspace}

\newcommand{\para}[1]{\vspace{2pt}{\bfseries{#1}}}

\newcolumntype{C}[1]{>{\centering\arraybackslash}p{#1}}

\def\BibTeX{{\rm B\kern-.05em{\sc i\kern-.025em b}\kern-.08em
    T\kern-.1667em\lower.7ex\hbox{E}\kern-.125emX}}
\begin{document}

\newcommand{\thiswork}{{LaZagna}}

\definecolor{highlighted}{rgb}{0.6,0.6,0.6}

\newcommand{\comment}[1]{\textcolor{orange}{General Comment: #1}}
\newcommand{\iy}[1]{\textcolor{blue}{IY:#1}}
\newcommand{\ch}[1]{\textcolor{red}{CH:#1}}

\title{\thiswork{}: An Open-Source Framework for Flexible 3D FPGA Architectural Exploration}

\author{
\IEEEauthorblockN{Ismael Youssef, Hang Yang, and Cong Hao}
\IEEEauthorblockA{School of Electrical and Computer Engineering, Georgia Institute of Technology \\
Email: \{ismael.youssef, hyang628, callie.hao\}@gatech.edu}
\vspace{-15pt}
}

\maketitle

\input{sections/0_Abstract}
\input{sections/1_Introduction}

\input{sections/2_Background_related}

\input{sections/3_Tool_flow}

\input{sections/4_Experiments}
\input{sections/5_Conclusion}

\bibliography{references}       
\bibliographystyle{IEEEtran}

\end{document}

%% file: sections/0_Abstract.tex
\begin{abstract}
While 3D IC technology has been extensively explored for ASICs, their application to FPGAs remains limited. Existing studies on 3D FPGAs are often constrained to fixed prototypes, narrow architectural templates, and simulation-only evaluations.
In this work, we present \thiswork{}, the first open-source framework for automated, end-to-end 3D FPGA architecture generation and evaluation. \thiswork{} supports high-level architectural specification, synthesizable RTL generation, and bitstream production, enabling comprehensive validation of 3D FPGA designs beyond simulation. It significantly broadens the design space compared to prior work by introducing customizable vertical interconnect patterns, novel 3D switch block designs, and support for heterogeneous logic layers. The framework also incorporates practical design constraints such as inter-layer via density and vertical interconnect delay.
We demonstrate the capabilities of \thiswork{} by generating synthesizable RTL that can be taken through full physical design flows for fabric generation, along with functionally correct bitstreams. Furthermore, we conduct five case studies that explore various architectural parameters and evaluate their impact on wirelength, critical path delay, and routing runtime. These studies showcase the framework’s scalability, flexibility, and effectiveness in guiding future 3D FPGA architectural and packaging decisions.
\thiswork{} is fully open-source and available on GitHub\footnote{\url{https://github.com/sharc-lab/LaZagna}} 
\end{abstract}

%% file: sections/1_Introduction.tex
\section{Introduction}

Field-Programmable Gate Arrays (FPGAs) are widely adopted in modern computing systems due to their flexibility, reconfigurability, and inherent parallelism. As application demands continue to grow in performance and integration complexity, improving the architectural efficiency of FPGAs has become a critical research direction.

Three-dimensional (3D) integration presents a promising path forward by vertically stacking logic, routing, and memory resources. In the ASIC domain, 3D integration has been extensively explored, demonstrating substantial gains in performance, density, and energy efficiency~\cite{Bamberg2020-iw, Nayak2015-si, Kim2021-ea}. In contrast, the application of 3D integration to FPGAs remains relatively underexplored, despite its potential to reduce interconnect delay, improve logic density, and support heterogeneous integration across layers.

Several prior works have proposed architectural enhancements for 3D FPGAs, such as separating configuration memory from logic~\cite{Turkyilmaz2014-vj, Lin2006-rm, Faaiq2025-qh}, or enabling multi-layer routing and homogeneous stacking~\cite{Le2009-qe, Amagasaki2015-nj, Amagasaki2018-ax, Boutros2023-og}. These studies report improvements in wirelength and critical path delay (CPD) compared to 2D counterparts. However, most proposals are limited to single or few designs, and rely solely on simulation-based evaluation without support for synthesizable RTL or bitstream generation. To advance the development of 3D FPGAs, there is a clear need for a powerful tool that can \textit{automatically and extensively explore the 3D FPGA architectural design space}. Such a tool would not only facilitate the discovery of optimal 3D FPGA architectures, but also provide actionable insights into viable 3D packaging technologies.

However, supporting comprehensive 3D FPGA design space exploration (DSE) poses multiple challenges. First, the tool must support a wide variety of vertical interconnect types and densities, to enable future studies on the physical integration of cross-layer vias. Second, heterogeneous layer configurations must be considered, including flexible partitioning of configurable logic blocks (CLBs), DSPs, and BRAMs across layers, aligning with emerging trends in 3D ICs. Third, new types of 3D switch blocks (SBs) and/or connection blocks (CBs) must be designed and evaluated. Fourth, vertical delay must be taken into account when organizing the layer structure. Finally, to enable practical validation, the framework must generate not only simulated outputs but also synthesizable RTL and functionally correct bitstreams for real benchmarks.

To address these challenges, we present \textbf{\thiswork{}}, the first open-source framework for \textit{automated 3D FPGA architecture generation and evaluation}. \thiswork{} enables rapid, end-to-end exploration of 3D FPGA designs, from high-level architectural specification to synthesizable RTL and bitstream generation. Unlike prior efforts focused on fixed designs or simulation-only flows, \thiswork{} offers comprehensive modeling of vertical connectivity, custom switch block design, heterogeneous logic layering, and seamless integration with open-source place-and-route and FPGA generation tools. We summarize our key contributions as follows:

\begin{enumerate}[leftmargin=*]
    \item \textbf{End-to-end and automated RTL and bitstream generation for 3D FPGAs:} \thiswork{} is the first framework to generate both synthesizable RTL and programming bitstreams for custom 3D FPGA fabrics, enabling evaluation beyond simulation. The tool is fully open-source to promote broad adoption and community-driven research.

    \item \textbf{Full-spectrum and comprehensive 3D architecture exploration:} \thiswork{} supports scalable exploration of a wide range of 3D architectural parameters, including layer count, resource partitioning, inter-layer connectivity, and routing granularity. Beyond supporting known architectures, it introduces a significantly broader design space, including novel vertical interconnect patterns, flexible 3D switch block designs, and logic heterogeneity across layers.

    \item \textbf{Evaluation and comprehensive case studies:} We demonstrate the capabilities of \thiswork{} by generating synthesizable RTL, performing full physical design flows, and producing functionally correct bitstreams for the generated 3D fabrics. To showcase the benefits of comprehensive DSE for future 3D FPGAs, we present five representative case studies that vary key architectural parameters and evaluate the designs in terms of wirelength, critical path delay, and routing runtime, demonstrating the framework's utility for informed architectural decision-making.
\end{enumerate}

%% file: sections/2_Background_related.tex
\section{Previous Work and Limitations}
\label{sec:background}

\input{figures/overview}
\input{tables/work_compare}

Related work falls into two main categories: (1) architectural studies on 3D FPGAs, and (2) tools for FPGA fabric and bitstream generation.

\para{3D FPGA Architecture Studies.}
Various strategies have been proposed for vertical integration and layer organization in 3D FPGAs. We categorize these architectures into the following classes:

\begin{itemize}[leftmargin=*]
    \item \textbf{Homogeneous:} All layers share an identical layout, including logic blocks (e.g., CLBs, DSPs, BRAMs), routing components (e.g., SBs, CBs), and configuration memory.
    
    \item \textbf{Non-Logic Heterogeneous:} Logic blocks are placed on one layer, while other components such as SBs, CBs, or configuration memory are located on separate layers.
    
    \item \textbf{Logic Heterogeneous:} Each layer contains a different combination of logic resources (e.g., CLBs, DSPs, BRAMs), while maintaining full routing capabilities across layers.
\end{itemize}

Table~\ref{tab:compare-with-existing} summarizes representative prior works on 3D FPGA architectures~\cite{Amagasaki2015-nj,Amagasaki2018-ax,Boutros2023-og,Turkyilmaz2014-vj,Faaiq2025-qh,Le2009-qe,Lin2007-pn}.
In the homogeneous category, Amagasaki et al.~\cite{Amagasaki2015-nj} evaluated a 3D design (type 1) with inter-layer CLB outputs, which yielded no CPD improvement. Their follow-up work~\cite{Amagasaki2018-ax} introduced a four-layer architecture with two logic and two routing layers but observed a 9\% degradation in CPD. Boutros et al.~\cite{Boutros2023-og} proposed a homogeneous design with vertical interconnects restricted to CLB and hard IP outputs, demonstrating gains of 4\% in wirelength and 3\% in CPD. Le et al.~\cite{Le2009-qe} investigated scalability across 2 to 10 layers and reported significant CPD improvements, up to 61\%, with a six-layer architecture.

Non-logic heterogeneous designs have also demonstrated great performance potential. Amagasaki et al.~\cite{Amagasaki2015-nj} also explored a type 2 architecture that separates logic and routing across layers using 3D switch blocks, achieving a 5.3\% CPD reduction over a 2D baseline. Turkyilmaz et al.~\cite{Turkyilmaz2014-vj} showed a 14\% CPD improvement by separating configuration memory from logic. Lin et al.~\cite{Lin2006-rm} isolated configuration memory onto a dedicated layer, resulting in an estimated 1.7$\times$ CPD improvement. Waqar et al.~\cite{Faaiq2025-qh} further advanced this idea using back-end-of-line integration to place configuration memory and SBs above CLBs, reporting CPD improvements ranging from 11\% to 30\%.

To the best of our knowledge, no prior work has discussed logic heterogeneous 3D FPGA architectures.

\para{FPGA Tools.}
For open-source FPGA architecture exploration, the most widely used tool is Verilog-to-Routing (VTR)~\cite{Murray2020-wj}, which supports placement and routing for customizable 2D FPGA fabrics. However, VTR's 3D capabilities are limited to two-layer homogeneous architectures, with inter-layer communication constrained to grid I/O pins. OpenFPGA~\cite{Tang2020-ac} builds on VTR and supports RTL and bitstream generation for user-defined architectures, but lacks support for 3D integration as well, posing a significant limitation for exploring vertically integrated designs.

\para{Limitations and \thiswork{} Capability.}
Despite notable progress in 3D FPGA research, several key limitations remain. Most prior studies are constrained to fixed or narrowly defined architectural templates, hindering systematic and generalizable design space exploration. Many omit critical components such as DSPs and BRAMs, which are essential in modern FPGA applications. Furthermore, existing evaluations are predominantly simulation-based and lack support for generating synthesizable RTL or programming bitstreams, making full end-to-end validation infeasible. No unified framework exists for consistent, apples-to-apples comparisons across different 3D integration strategies, and important architectural dimensions, such as via placement, layer heterogeneity, and switch block topology, have not been explored in a cohesive manner.

In contrast, \thiswork{} fills this gap by providing the first fully automated, open-source framework for comprehensive 3D FPGA architecture generation and evaluation. It unifies architectural modeling, RTL and bitstream generation, and performance evaluation into a scalable infrastructure, enabling systematic exploration across a broad range of 3D design spaces.

%% file: figures/overview.tex
\begin{figure*}[t] 
    \centering
    \includegraphics[width=\textwidth,keepaspectratio=true]{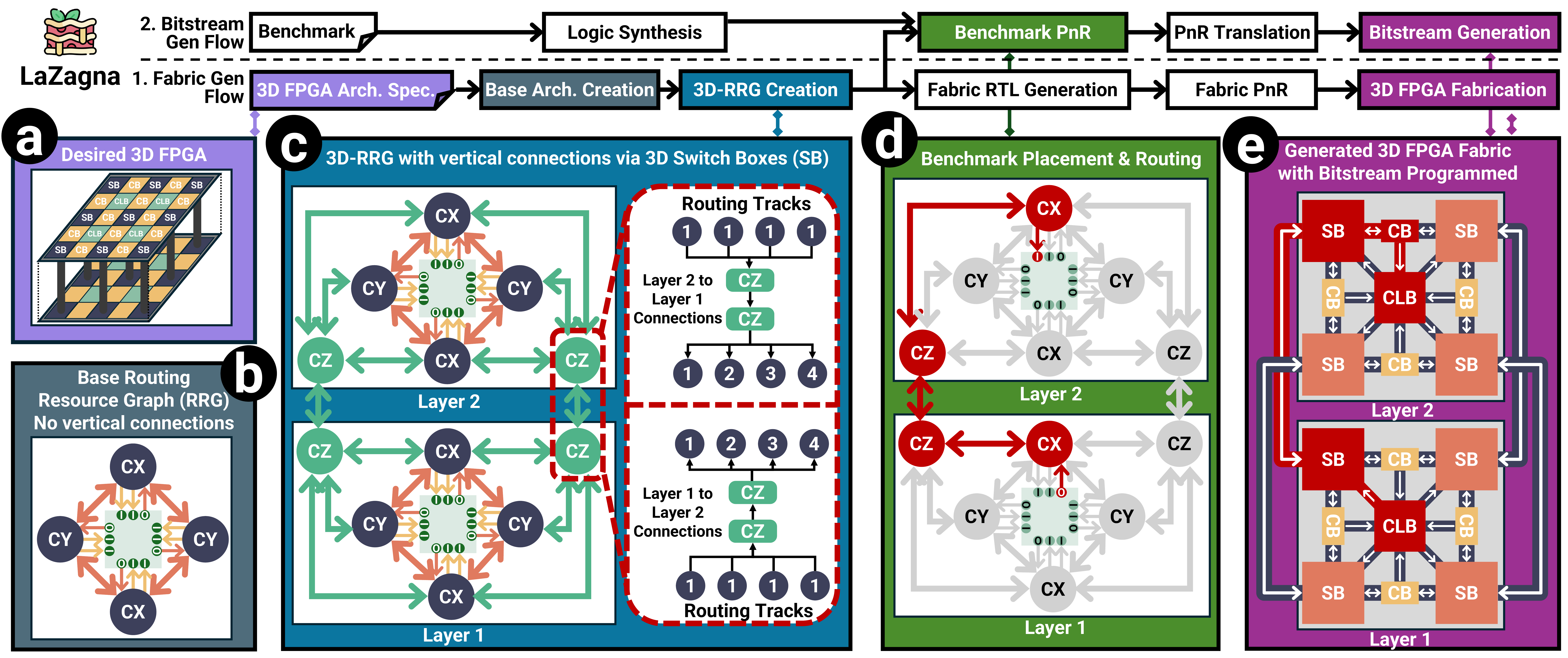}
    \caption{\thiswork{}  overview. It has two parallel flows: 1. 3D FPGA fabric RTL generation 2. benchmark bitstream generation. }
    \label{fig:tool_workflow}
\end{figure*}

%% file: tables/work_compare.tex
\begin{table*}[t]
    \centering

    \begingroup
    \scriptsize
    \setlength\tabcolsep{2.8pt}

    \begin{tabular}{l | c | c | c | c | c | c | >{\columncolor{cyan!15}}c }
        \toprule
        \textbf{FPGA Architecture} & \textbf{\cite{Amagasaki2015-nj} type 1} & \textbf{\cite{Boutros2023-og}} & \textbf{\cite{Le2009-qe}} & \textbf{\cite{Amagasaki2015-nj} type 2} & \textbf{\cite{Amagasaki2018-ax}} & \textbf{\cite{Turkyilmaz2014-vj}, \cite{Lin2007-pn}, \cite{Faaiq2025-qh}} & \textbf{\thiswork{} Options} \\
        \hline
        \textbf{\# of Layers} & 2 & 2 & 2--10 & 2 & 4 & 2 & Arbitrary \\
        \hline
        \textbf{Layer Type (Sec. \ref{sec:parameters})} & Homo & Homo & Homo & Non-Logic Hetero & Non-Logic Hetero & Non-Logic Hetero & Logic Hetero \\
        \hline
        \textbf{Vertical Connection Type} & 3D CB & 3D CB-O & 3D SB & 3D SB & 3D CB-O & -- & \cellcolor{cyan!15}Self-defined (Table~\ref{tab:arch_parameters}) \\
        \hline
        \textbf{Layers Configuration} & -- & -- & -- & 1 Routing, 1 Logic & 2 Routing, 2 Logic & 1 Layer for Config. Memory & \cellcolor{cyan!15}Self-defined (Table~\ref{tab:arch_parameters}) \\
        
        \midrule

        \rowcolor{cyan!15}
        \multicolumn{8}{c}{\textbf{\thiswork{}} supports \textbf{all} features listed above and more (summarized in Table~\ref{tab:arch_parameters}), including FPGA architectural RTL and benchmark bitstream generation.} \\

        \bottomrule
    \end{tabular}
    \endgroup
    \caption{\thiswork{}'s 3D FPGA modeling capabilities compared to previous works.}
    \label{tab:compare-with-existing}
    \vspace{-8pt}
\end{table*}

%% file: sections/3_Tool_flow.tex
\section{Tool Workflow}
\label{sec:tool_wrokflow}

Figure~\ref{fig:tool_workflow} illustrates the overall workflow of \thiswork{}, which consists of two distinct flows: one for generating the FPGA fabric and another for generating the bitstream used in benchmark evaluation.

\textbf{1. Fabric Generation Flow:} This flow begins with a user-defined description of the target 3D FPGA architecture, specified via configurable input parameters. Based on this description, \thiswork{} generates synthesizable RTL using a customized version of OpenFPGA with extended support for 3D integration. The RTL can then be passed through standard physical design steps, including placement and routing (PnR)—a process we refer to as \textbf{fabric PnR}—to produce a GDSII layout ready for fabrication, and accurate PPA metrics.

\textbf{2. Bitstream Generation Flow:} In this flow, the user provides benchmark circuits to evaluate the performance of the generated 3D FPGA. These circuits are mapped onto the architecture using VTR, in a process we call \textbf{benchmark PnR}, to clearly distinguish it from fabric PnR. After benchmark PnR, key performance metrics such as WL and CPD are extracted. The routing results are then passed to our customized 3D-enabled OpenFPGA to generate a programming bitstream, which is used to configure the 3D FPGA.

\input{tables/arch_parameters}
\input{figures/tool_parameters}

\subsection{3D FPGA Architectural Parameters}
\label{sec:parameters}

\thiswork{} accepts several architectural parameters that allow users to define custom 3D FPGA configurations, as illustrated in Fig.~\ref{fig:tool_workflow}(a).  
Table~\ref{tab:arch_parameters} summarizes all supported parameters, including both standard options and new configurations introduced by \thiswork{}.  
Fig.~\ref{fig:tool_parameters} visualizes some parameters.  
This section details the key parameters supported by \thiswork{}.

\subsubsection{Vertical Connection Types (Fig.~\ref{fig:tool_parameters_a})}  

Vertical connections are one of the most critical considerations in 3D FPGA design, as they determine how layers are interconnected and directly impact WL, CPD, and vertical via count and density.  
As shown in Table~\ref{tab:compare-with-existing}, previous studies explore only a limited subset of these options. \thiswork{} not only supports all those types but also introduces new, highly configurable vertical interconnect schemes.

We categorize vertical connection types into the following classes, visualized in Fig.~\ref{fig:tool_parameters_a}. For clarity, only connections from Layer 2 to Layer 1 are shown; the reverse direction is omitted but implied.

\begin{itemize}[leftmargin=*]
    \item \textbf{3D CB:} All input and output pins of grid elements (including CLBs, IOs, DSPs, and BRAMs) are allowed to connect across layers via vertical vias. This configuration enables layer crossings at both the source and destination of a net and is natively supported by VTR. As shown in Fig.~\ref{fig:tool_parameters_a}, CBs in Layer 2 can connect to CLBs in Layer 1, and vice versa. While this approach offers maximum routing flexibility, it incurs a high vertical via count.

    \item \textbf{3D CB-O:} To reduce via usage, this variant restricts cross-layer connectivity to only the output pins of grid elements. Thus, nets only traverse layers at their source. This is also supported by VTR.

    \item \textbf{3D SB:} Cross-layer connections are confined to switch blocks. Nets may change layers only when encountering a designated 3D SB. This model localizes vertical vias to routing junctions, potentially simplifying integration and enabling more controlled via placement.

    \item \textbf{3D Hybrid:} This configuration combines the benefits of both 3D CB and 3D SB designs by allowing cross-layer input/output pin connections and SB-based crossings. It maximizes routing flexibility and vertical utilization but leads to a higher via count.

    \item \textbf{3D Hybrid-O:} A variant of 3D Hybrid that restricts cross-layer connections to output pins only. It provides greater flexibility than 3D CB-O or 3D SB alone while reducing vertical via demand compared to the full 3D Hybrid model.

    \item \textbf{Others (User-Defined):}  
    Beyond the predefined types, \thiswork{} allows users to define custom vertical connection strategies. For instance, \emph{3D CB-I} or \emph{3D Hybrid-I} can be specified to enable only input pin crossings. Users may also configure partial inter-layer connectivity, such as selecting specific pins or SB tracks for vertical connections. This fine-grained control enables exploration of a significantly broader and more realistic 3D design space.
\end{itemize}

Each vertical connection type presents trade-offs between routing flexibility and vertical interconnect overhead, making it a critical design knob for future 3D FPGA architecture DSE. 
We will conduct a detailed case study comparing these connection types in Sec.~\ref{sec:3D-connection-type-study}.

\subsubsection{3D SB Placement Locations (Fig.~\ref{fig:tool_parameters_e})} 

\thiswork{} allows users to configure both the percentage and spatial distribution of 3D SBs across the FPGA grid. Several default placement patterns are supported, including: \textbf{Repeated Interval}, \textbf{Rows}, \textbf{Columns}, \textbf{Core}, \textbf{Perimeter}, and \textbf{Random}. Users may also define custom placement strategies.
For each pattern, users can specify the desired percentage of SBs to be 3D, and \thiswork{} automatically calculates the corresponding number and distributes them accordingly across the fabric. Additionally, users may provide an explicit list of 3D SB locations via a CSV file, allowing full manual control over 3D SB placement. The impact of various 3D SB placement strategies will be evaluated through a dedicated case study in Section~\ref{sec:case_study_sb_palcement}.

\subsubsection{3D Switch Block Connection Patterns (Fig.~\ref{fig:tool_parameters_b})}
\label{sec:3D-SB}

To extend a traditional 2D SB to 3D, vertical cross-layer routing tracks are added. While a 2D SB connects tracks from four planar sides—left, bottom, right, and top—a 3D SB can support up to six sides by including vertical connections to layers above and below.

As in 2D FPGAs, the routing behavior of a switch block is governed by a connection pattern that defines how tracks from different sides are interconnected. In the 3D context, this pattern must also account for vertical routing. Two types of patterns must be specified: the output pattern, which determines which planar tracks drive each cross-layer output track, and the input pattern, which determines how each cross-layer input track connects to planar tracks.

In \thiswork{}, these patterns are user-defined. Both the input and output patterns are expressed as a sequence of four integers corresponding to the four planar sides of the switch block in counter-clockwise order. Each integer indicates the starting track index on that side to be used in vertical connectivity.
Consider a switch block with channel width $W$ and an output pattern defined as $[i_0, i_1, i_2, i_3]$. The $k$-th output cross-layer track will be driven by the planar tracks indexed as $(i_0 + k) \bmod W$ on the left side, $(i_1 + k) \bmod W$ on the bottom, $(i_2 + k) \bmod W$ on the right, and $(i_3 + k) \bmod W$ on the top, where $k$ ranges from $0$ to $W-1$. The input pattern operates analogously, except that the specified planar track indices are the targets driven by each vertical input track.

Fig.~\ref{fig:tool_parameters_b} presents a variety of input and output patterns and visualizes how they are realized within the 3D SB structure. In FPGA architectures with segment lengths greater than one, it is possible for the number of input tracks to exceed the number of output tracks. In such cases, the defined pattern is repeated until all input tracks are assigned an output.

\subsubsection{Vertical Connection Delay Ratio (Fig.~\ref{fig:tool_parameters_c})} 

This parameter specifies the delay cost of vertical interconnects relative to horizontal connections. Since \thiswork{} is designed to be technology-independent, users can define a delay ratio that scales the vertical connection delay with respect to a baseline, like the delay of a driver switch on the fabric. 
Alternatively, users may specify the absolute delay value for vertical interconnects if desired.
This abstraction enables exploration of different vertical delays without relying on a technology node.

\subsubsection{Layer Count and Heterogeneity (Fig.~\ref{fig:tool_parameters_d})} 
\label{sec:parameters_layer_count_and_heterogenity}

This parameter allows users to specify both the number of layers in the 3D FPGA and the type of blocks, such as logic elements and connection components, assigned to each layer. As discussed in Section~\ref{sec:background}, \thiswork{} supports three types of layer configurations: \textbf{Homogeneous}, \textbf{Non-Logic Heterogeneous}, and \textbf{Logic Heterogeneous}. 
For example, for a 3-layer logic heterogeneous architecture, a user may choose to place DSP blocks only on the top layer to improve thermal dissipation, allocate additional BRAM blocks to a middle layer for memory balancing, and concentrate routing elements such as SBs and CBs on the bottom layer.

\subsubsection{All Standard 2D FPGA Parameters}

\thiswork{} also supports all conventional parameters used in 2D FPGA architectures, as provided by VTR and OpenFPGA. These include grid size, channel width, LUT size, inclusion and configuration of hard IP blocks (e.g., DSPs and BRAMs), and others.

\para{Vast design space.}
With all these configurable parameters and support for easy future extensions, \thiswork{} enables exploration of a vast and flexible design space for 3D FPGAs. 
For example, even when 2D FPGA parameters are fixed, only two homogeneous layers are used, the input/output pattern indices for 3D SBs are restricted to the range of -3 to 3, and only the default 3D SB placement patterns are considered, there are still \textbf{1,729,440,302 unique 3D FPGA configurations} possible by changing the Connection Type and SB placement, and pattern. All of these configurations can be placed and routed, and have corresponding RTL and bitstreams generated using \thiswork{}.
When custom 3D SB placements are also included, the design space grows exponentially—exceeding 2\textsuperscript{130} possible configurations—even without considering any 2D FPGA parameter variations. While many of these configurations may share similar characteristics, this immense design space highlights the potential of \thiswork{} to drive the exploration of future 3D FPGA architectures.

\subsection{3D Fabric Routing Resource Graph Creation}
\label{sec:rrg}

Once the 3D FPGA architectural parameters are specified and parsed, \thiswork{} constructs a base Routing Resource Graph (RRG) (Fig.~\ref{fig:tool_workflow}b) and extends it to a 3D-RRG (Fig.~\ref{fig:tool_workflow}c). This RRG is essential for both fabric RTL generation and bitstream creation.

\para{Base RRG.}
The base RRG represents the FPGA’s interconnect structure and is utilized by VTR during benchmark placement and routing (PnR). It serves two key roles. First, it allows benchmark PnR to be formulated as an optimization problem that aims to minimize the critical path delay (CPD) while satisfying architectural and placement constraints. Second, because the RRG encodes the complete connectivity of the fabric, it is also used during RTL generation in OpenFPGA to classify each node and edge by its corresponding architectural block (e.g., CLB, SB, etc.).

As shown in Fig.~\ref{fig:tool_workflow}(b), the RRG includes six standard node types defined by VTR: {IPIN}, {OPIN}, {SOURCE}, {SINK}, {CHANX}, and {CHANY}. {IPIN} and {OPIN} (labeled 'I' and 'O' in the figure) represent the input and output pins of logic blocks at each grid location. {SOURCE} and {SINK} (not shown in the figure) interface between the routing network and internal logic, where each {SINK} connects to an {IPIN}, and each {SOURCE} connects to an {OPIN}. {CHANX} and {CHANY} ('CX' and 'CY') represent horizontal and vertical routing tracks, respectively, within the same layer.

For 3D architectures that do not use 3D switch blocks, for example 3D CB and 3D CB-O, the base RRG already includes all required connectivity, as these designs are natively supported by VTR.

\para{3D-RRG.}
To support architectures with custom 3D SB designs, \thiswork{} extends the base RRG to a 3D-RRG by introducing additional nodes and edges, as illustrated in Fig.~\ref{fig:tool_workflow}(c).

As discussed in Section~\ref{sec:3D-SB}, in a physical 3D SB, each input track on the source layer is selected via a multiplexer based on the input pattern. The selected signal is then transmitted to the sink layer and distributed to output tracks according to the output pattern.

To represent this in the 3D-RRG, \thiswork{} introduces a new node type: {CHANZ} ('CZ' in figure), which models vertical routing. For each inter-layer SB connection, one {CHANZ} node is created on the source layer and one on the sink layer. The source-layer {CHANZ} node connects input tracks to the sink-layer {CHANZ} node, based on the SB’s input pattern. The sink-layer node then fans out the signal to output tracks according to the output pattern. 

To describe the direction of vertical connections, \thiswork{} extends VTR’s standard routing directions, {Incrementing} and {Decrementing} for planar routing, and introduces four new vertical directions: \textbf{Above Incrementing}, \textbf{Above Decrementing}, \textbf{Under Incrementing}, and \textbf{Under Decrementing}. \textit{Above} or \textit{Under} denotes the side of the layer the connection is on. \textit{incrementing} or \textit{decrementing} refers to the track's direction between the layers. \textit{Incrementing} means the signal is going from a lower layer to a higher one, \textit{decrementing} means the opposite. 

For example, a connection labeled \textit{Above Incrementing} indicates that the signal travels upward (i.e., towards a higher layer index), this implies that it is an output of the current layer. Although these directions are not utilized by VTR’s algorithms, they are useful for RTL generation, as they explicitly encode the vertical connectivity structure. This allows \thiswork{} to efficiently identify source and sink layers for each inter-layer connection during RTL generation.
\vspace{-2pt}
\subsection{3D FPGA Architecture RTL Generation}
\label{sec:rtl_generation}

Once the 3D-RRG is constructed, \thiswork{} modifies the OpenFPGA flow to generate synthesizable RTL for the 3D FPGA architecture. The primary enhancement involves recognizing and handling the new node and edge types introduced in the 3D-RRG.

\para{3D-RRG Annotation.}
Annotation refers to mapping nodes and edges in the RRG to their corresponding physical components in the FPGA architecture (e.g., CLBs, SBs). To correctly generate a 3D fabric, our modified OpenFPGA annotates all elements of the RRG, including newly added inter-layer connections, with their associated logic or routing blocks. Specifically, \thiswork{} annotates the following cross-layer signals:

\begin{itemize}[leftmargin=*]
    \item \textbf{Inter-layer Grid Inputs:} Each CB must identify inter-layer routing tracks and connect them to the correct input pins of grid locations. To manage this, each CB performs two tasks: inter-layer output identification and input reception. First, during its creation, each CB analyzes its routing tracks' outgoing edges to identify inter-layer edges and outputs them. In the second step, the CB checks if another CB at the same location on a different layer is sending an inter-layer signal. If such a signal exists, the inter-layer connection is added as a connection to the CB, and is connected to the correct pin's input multiplexer. 

    \item \textbf{Inter-layer Grid Outputs:} Each SB is responsible for receiving inter-layer grid output signals and connecting them to the appropriate routing tracks on the SB's layer. To manage this, each SB, during its creation, checks for connections from other layers' grid location outputs. If such a connection exists, its added to the SB and connected to the respective multiplexers of the routing tracks to which it links.

    \item \textbf{3D SB Signals.} Each 3D SB must also correctly identify inter-layer signals from other 3D SBs. To send an inter-layer signal, the SB multiplexes input routing tracks based on the SB pattern, marking the output as an inter-layer signal. To receive an inter-layer signal, the SB inspects other SBs with the same location but on different layers; if an inter-layer output is detected, the SB de-multiplexes it based on its output pattern and routes it to the appropriate output tracks.
\end{itemize}

\para{RTL Generation.}
Once annotation is complete, RTL generation becomes a deterministic mapping process. A standard cell library (including flip-flops, multiplexers, LUTs, etc.) is provided to OpenFPGA. Using this library, the fabric RTL is constructed by translating the annotated RRG into a complete netlist. Each connection and logic block from the graph is assigned a corresponding RTL primitive, and grouped into higher-level modules.

The generated RTL is organized hierarchically to reflect the layered structure of the fabric. For example, in a two-layer architecture, the output includes files such as \texttt{top.v}, \texttt{layer1.v}, and \texttt{layer2.v}. Each layer file contains only the modules relevant to that layer (e.g., CLBs, SBs).
These modules continue to decompose hierarchically down to individual standard cell instances. All cross-layer signals are explicitly defined in the top module, making inter-layer connectivity easy to trace.

Importantly, this RTL generation process is independent of the benchmark PnR. If the 3D-RRG can be created, the RTL can be generated, even for fabrics not currently supported by VTR. This allows \thiswork{} to produce RTL for logic-heterogeneous 3D architectures and those with more than two layers, which are beyond the capabilities of VTR.

For non-logic heterogeneous architectures, the resulting RTL is structurally similar to that of a 2D FPGA. The only difference lies in module hierarchy: modules are organized by physical layer rather than as a single design. This separation improves design clarity and facilitates physical layout by clearly delineating the blocks assigned to each layer.

\input{figures/fpga_implementation}

\subsection{Benchmark Logic Synthesis, Placement, and Routing}
\label{sec:benchmark_pnr}

With the 3D-RRG generated, the benchmark circuit undergoes logic synthesis with Yosys (performed independently of 3D-RRG), followed by benchmark PnR onto the 3D fabric using VTR. 
Since the 3D-RRG reflects the architecture of the target FPGA, VTR can apply its PnR algorithms directly and produce valid routing results. These results include performance metrics such as CPD, WL, and resource utilization.
We note that a key limitation of VTR's 3D placement algorithm is that logic blocks are assigned to specific layers during initial placement and are rarely allowed to move across layers thereafter. This constraint may lead to suboptimal placements. Moreover, VTR does not currently support heterogeneous 3D architectures or CB-based vertical connectivity for architectures with more than two layers. Addressing these limitations remains an avenue for future toolchain enhancements.

\subsection{Bitstream Generation}

Following benchmark placement and routing, \thiswork{} extends OpenFPGA to generate the programming bitstream for the 3D fabric. 
The process mirrors the RRG annotation procedure described in Section~\ref{sec:rtl_generation}, but instead of annotating all nodes and edges in the RRG, it focuses solely on the routed nodes and edges reported by VTR. As shown in Fig.~\ref{fig:tool_workflow}(d) where routed edges are highlighted in red. OpenFPGA traces each signal path, determines the sequence of blocks traversed, and identifies the corresponding configuration bits to activate the routing resources.
For 3D grid inputs, this involves configuring CB multiplexers to correctly transmit and receive inter-layer signals. For 3D grid outputs and 3D SBs, the configuration memory bits of the corresponding multiplexers are updated to reflect vertical connections. These enhancements ensure that the generated bitstream correctly configures the 3D FPGA to match the routing determined during benchmark PnR.

\subsection{Tool Output}

\thiswork{} produces three primary outputs: the RTL for the 3D FPGA fabric, benchmark PnR results, and the corresponding bitstream.
The RTL enables downstream analysis and physical design, allowing designers detailed power, performance, area, and thermal (PPAT) evaluations, and potentially GDSII layouts for 3D FPGAs.

%% file: tables/arch_parameters.tex
\begin{table}[tb]
    \centering
    \scriptsize
      \setlength\tabcolsep{3pt}
    \begin{tabular}{C{0.25cm}|C{2.5cm}|C{5cm}}
        \toprule
        \rowcolor{gray!25}
         & \textbf{FPGA Parameters} & \textbf{Options} \\
        \hline
        (1) & Vertical Connection Types (Fig.~\ref{fig:tool_parameters_a}) & 3D CB, 3D CB-O, 3D SB, 3D Hybrid, 3D Hybrid-O (can be easily extended) \\ \hline

        (2) & 3D SB Percentage and Locations (Fig.~\ref{fig:tool_parameters_e}) & 0 to 100\% with placement pattern: [Perimeter, Center, Random, Repeated Interval, Custom] \\ \hline
        
        (3) & 3D Switch Block Patterns (Fig.~\ref{fig:tool_parameters_b}) & Any two 4-integer combinations, e.g., Input [0,1,2,3], Output [1,2,3,4] \\\hline
        
        (4) & Vertical Connection Delay Ratio (Fig.~\ref{fig:tool_parameters_c}) & A floating value describing the delay ratio between horizontal and vertical connections \\ \hline
        
        (5) & Layer Count and Heterogeneity (Fig.~\ref{fig:tool_parameters_d}) & Arbitrary layer count; layers can be: Homo, Non-Logic Hetero, Logic Hetero \\ \hline

        (6) & Normal FPGA Parameters & Channel Width, Grid Size, LUT Size, Segment Length, etc. \\

        \bottomrule
        
    \end{tabular}
    \caption{Supported 3D FPGA architectural parameters.}
    \label{tab:arch_parameters}
\end{table}

%% file: figures/tool_parameters.tex
\begin{figure}[t]
    \centering
    \begin{subfigure}{\columnwidth}
        \centering
        \includegraphics[width=0.95\linewidth]{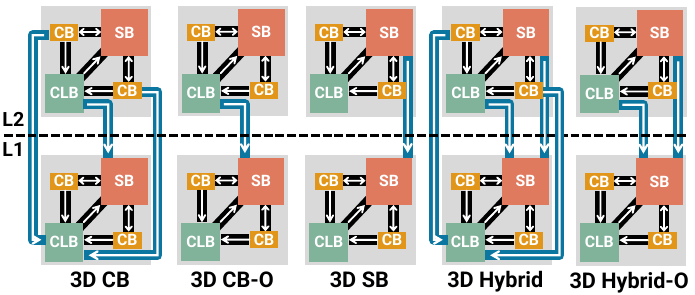}
        \caption{Supported different vertical connection types.}
        \label{fig:tool_parameters_a}
    \end{subfigure}
    
    \vskip\baselineskip

    \begin{subfigure}{\columnwidth}
        \centering
        \includegraphics[width=\linewidth]{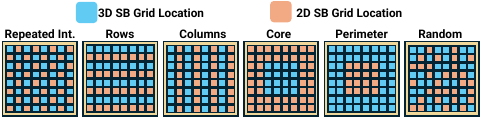}
        \caption{Examples of different 3D SB placement strategies.}
        \label{fig:tool_parameters_e}
    \end{subfigure}
    
    \vskip\baselineskip

    \begin{subfigure}{\columnwidth}
        \centering
        \includegraphics[width=0.95\linewidth]{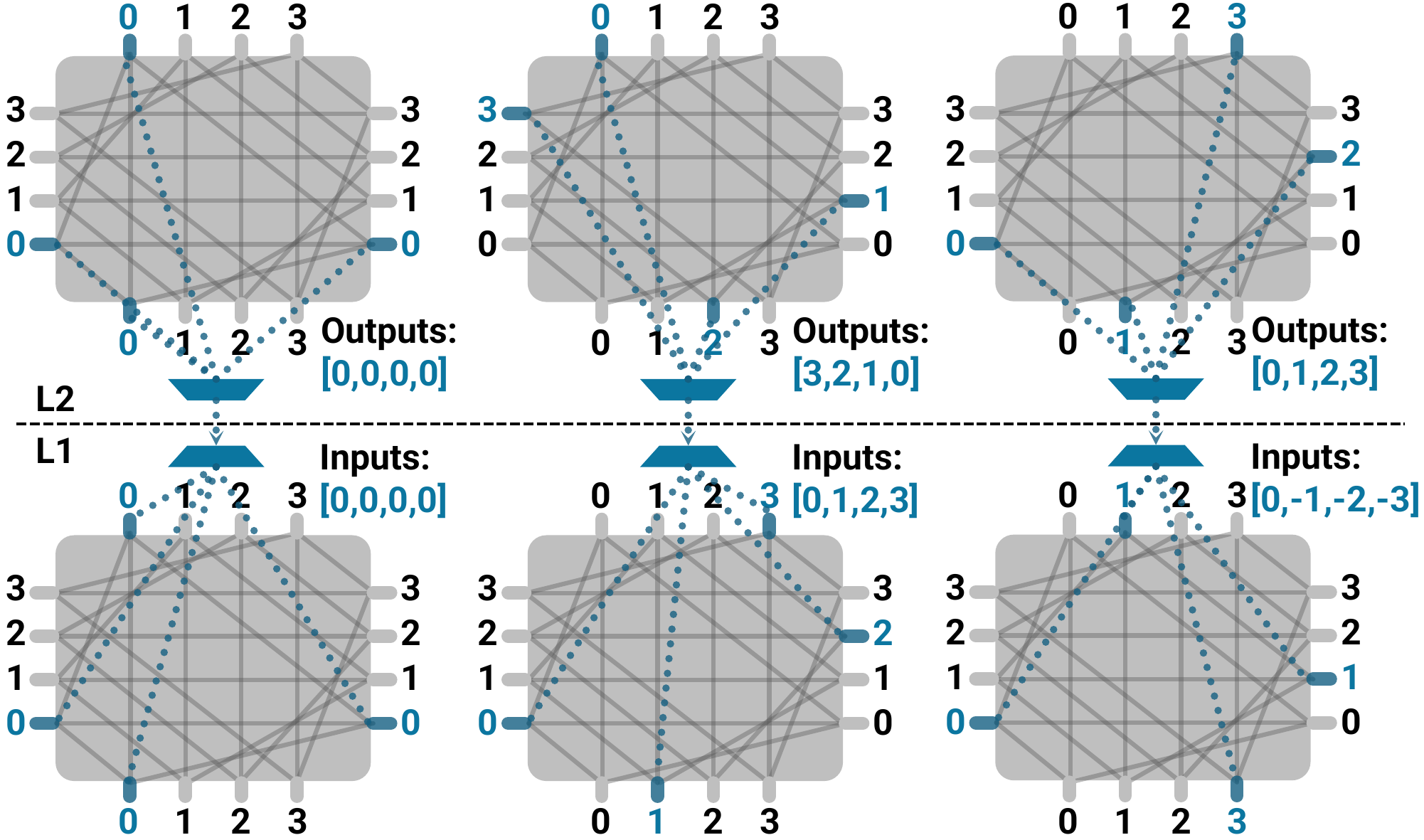}        \caption{Examples of different 3D SB connection patterns.}
        \label{fig:tool_parameters_b}
    \end{subfigure}
    
    \vskip\baselineskip
    
    \begin{subfigure}{0.65\columnwidth}
        \vspace*{0pt} 
        \centering
        \includegraphics[width=\linewidth]{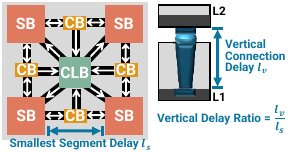}
        \caption{Vertical delay ratio.}
        \label{fig:tool_parameters_c}
    \end{subfigure}
    \begin{subfigure}{0.28\columnwidth}
        \vspace*{0pt} 
        \centering
        \includegraphics[width=\linewidth]{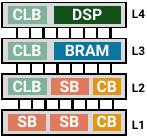}
        \caption{Layer heterogeneity.}
        \label{fig:tool_parameters_d}
    \end{subfigure}

    \caption{\thiswork{} input parameters.}
    \label{fig:tool_parameters}
    \vspace{-2pt}
\end{figure}

%% file: figures/fpga_implementation.tex
\begin{figure}[bt]
    \centering
    \includegraphics[width=0.9\columnwidth]{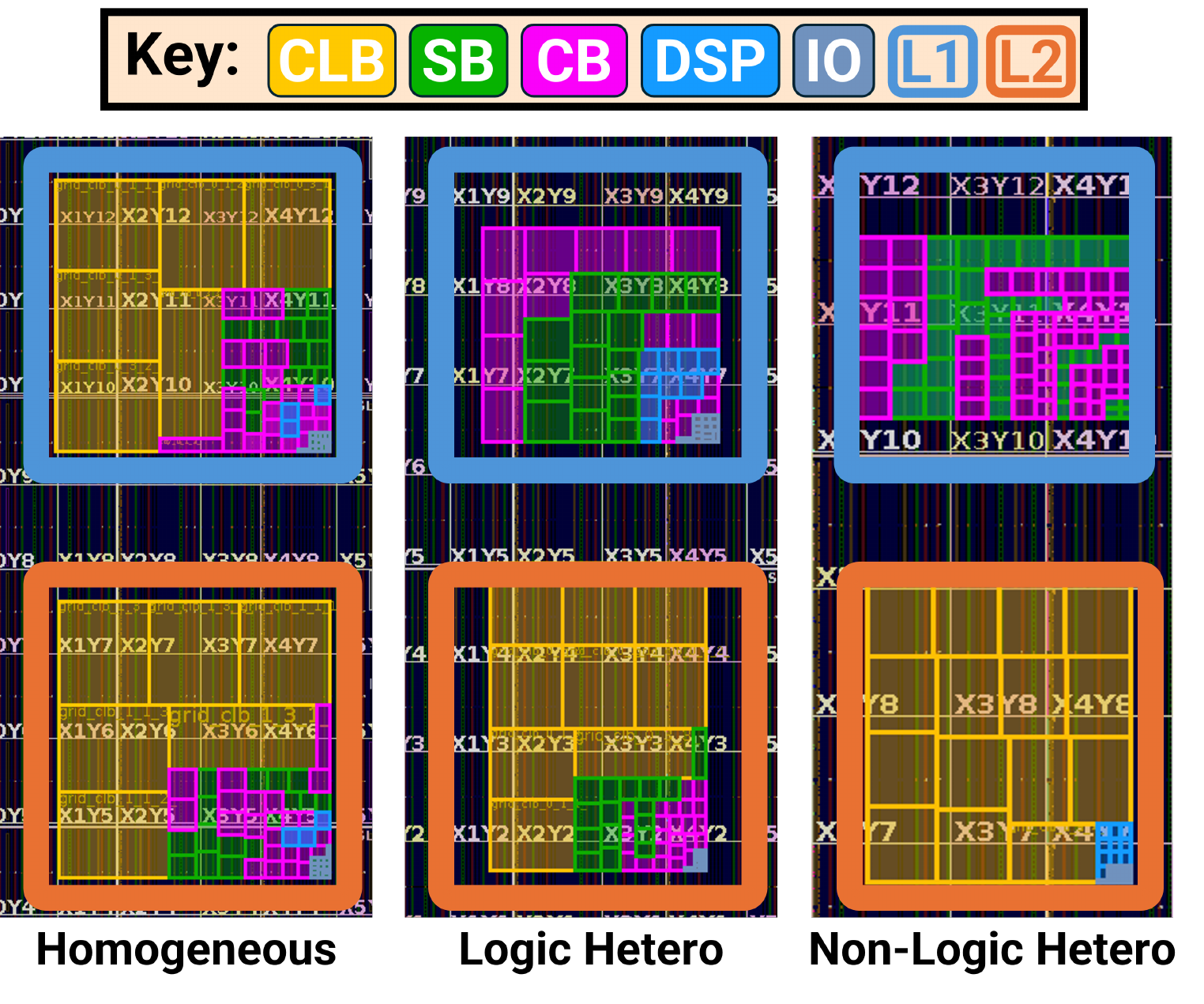}
    \caption{Implementation of different layer heterogeneity 3D FPGAs onto a Xilinx Virtex Ultrascale+ using RTL generated by \thiswork{}}
    \label{fig:fpga_implementation}
\end{figure}

%% file: sections/4_Experiments.tex
\section{Experimental Results}

\input{figures/physical_implementation}

This section demonstrates the capabilities of \thiswork{} in two critical aspects. First, we show that synthesizable RTL for 3D FPGA fabrics can be automatically generated, producing valid bitstreams and maintaining compatibility with industry-standard physical design tools. Second, to highlight the framework’s utility for architectural exploration, we conduct five case studies by varying key 3D FPGA design parameters, such as vertical interconnect types and 3D SB designs, and evaluate key performance metrics. Importantly, this work does not advocate for a specific 3D architecture nor claims superiority over conventional 2D designs, as performance is inherently dependent on packaging and process technologies. Rather, our goal is to enable systematic and reproducible 3D FPGA design space exploration.

\subsection{Experimental Setup}

We evaluate the generated 3D FPGA fabrics using a subset of the Koios benchmark suite \cite{koios}, as also adopted in prior work \cite{Boutros2023-og}. Koios provides a representative collection of large-scale, contemporary RTL designs that heavily utilize embedded DSPs and BRAMs—features critical to realistic architectural assessments.
We selected a subset of Koios benchmarks that satisfy two criteria: (1) they can be placed and routed on a 2D baseline FPGA fabric within one hour, and (2) they fit within the logic and memory resources of our target grid. 

All experiments were conducted on a server running Red Hat Enterprise Linux 8.5.0 with a Xeon 6226R processor and 512 GB of RAM. PnR of the benchmarks were done on VTR version 8.1. Fabric RTL and bitstream generation were done using our customized OpenFPGA 1.2 flow.

\subsection{Tool Demonstration}

\para{1. RTL and Fabric Generation} To demonstrate the RTL generation capabilities of \thiswork{}, three 2-layered 3D FPGA fabrics with varying layer heterogeneity were designed. These designs were synthesized and implemented on a Xilinx Ultrascale+ FPGA using Vivado. The resulting floorplans are shown in Figure \ref{fig:fpga_implementation}. The types of heterogeneity tested were: homogeneous, where the layers are the same; logic heterogeneous, with DSPs on one layer and CLBs on the other; and non-logic heterogeneous, with SBs and CBs on one layer, and DSPs and CLBs on the other. 

To verify both the functionality of the generated 3D FPGA RTL and the validity of the bitstream generated, the s298 MCNC benchmark \cite{mcnc} was implemented onto the homogeneous and non-logic heterogeneous fabrics. Bitstream generation is not currently possible for the logic heterogeneous architectures, as explained in Section \ref{sec:benchmark_pnr}. The bitstream was generated using the flow described in Section \ref{sec:tool_wrokflow}. The fabrics were then functionally verified by programming them with the bitstream in RTL simulation and comparing their outputs to the expected results. This process confirmed the correctness of both the generated RTL and bitstream.

\para{2. Physical Implementation} To validate the fabric PnR feasibility of the generated RTL, we physically implement a 4×4 homogeneous 3D-SB FPGA architecture using Cadence\textregistered~Genus and Innovus. The implementation leverages the open source FreePDK45 process \cite{Stine2007-ug} with a modified Routing Technology Kit (RTK) LEF, upgraded to version 5.8 with face-to-face stacking, to enable compatibility with the Innovus 3D flow. The congestion map and routing result of the fabric are shown in Fig.\ref{fig:pdResult}. This demonstrates that our fabric RTL can be used with commercial physical design tools to achieve accurate PPA and fabricate 3D FPGAs.

\para{3. Tool Runtime} Fig.~\ref{fig:runtime_results} presents the end-to-end runtime of \thiswork{}, using the \texttt{eltwise\_layer} benchmark across various FPGA grid sizes. As expected, 3D FPGA configurations incur higher runtime than their 2D counterparts with equivalent grid dimensions. This increase is primarily due to the added complexity of 3D fabrics, which feature a greater number of interconnect edges and logic nodes that must be processed during the tool workflow.
Importantly, the runtime exhibits a near-linear relationship with the number of grid locations, demonstrating that \thiswork{} scales predictably with fabric size, making it feasible to apply the tool to larger architectures without incurring prohibitive overhead.

\input{figures/runtime_analysis}
\input{tables/case_studies_parameters}

\subsection{Case Studies}
\label{sec:case_studies}

To demonstrate the capabilities of \thiswork{}, we conducted five case studies to explore the 3D FPGA architectural design space. These studies highlight the tool’s flexibility and effectiveness in enabling comprehensive architectural exploration.
Table~\ref{tab:case_studies_paramters} summarizes the 3D FPGA parameters used across the case studies, indicating which parameters remain fixed and which are varied in each specific study.

Key evaluation metrics for 3D FPGA architectures include the number of required vertical interconnects, total wirelength (\textbf{WL}), and critical path delay (\textbf{CPD}). It is important to note that both WL and CPD are highly sensitive to the quality of PnR results, physical design technology node, and 3D packaging technology. Consequently, the reported WL and CPD values may vary significantly with improved 3D-aware PnR tools or the adoption of advanced packaging solutions.

\subsubsection{Impact of Vertical Connection Types}
\label{sec:3D-connection-type-study}

In this study, we compare a variety of 3D connection types against a baseline 2D architecture to evaluate the aforementioned metrics. Each architecture is configured to have a comparable number of CLBs, DSPs, and BRAMs, while varying the routing resources.

Specifically, we evaluate the following vertical connection types: {3D CB}, {3D CB-O}, {3D SB}, {3D Hybrid}, and {3D Hybrid-O}, as defined in Section~\ref{sec:parameters} and illustrated in Fig.~\ref{fig:tool_parameters_a}. For architectures utilizing 3D SBs, we vary the proportion of SBs with vertical connections to be 100\% (all SBs are 3D), 66\% (two-thirds), and 33\% (one-third).

For the Hybrid configurations, this percentage applies only to the SBs on the grid that incorporate 3D connectivity. In {3D Hybrid} architectures, all input and output pins of grid locations have inter-layer connections, and only the output pins are interlayer for {3D Hybrid-O} architectures.

\para{Number of vertical connections.}
3D FPGA design is fundamentally constrained by the density of available vertical inter-layer connections, which is determined by the chosen integration technology (e.g., TSVs, MIVs, or microbumps). For instance, hybrid bonding offers up to 302 vertical connections per grid location at a 1 µm pitch, but only 12 connections at a 5 µm pitch, based on the ASAP7 PDK \cite{Boutros2023-og}. MIVs provide a more attractive solution for 3D FPGAs due to their finer pitch, down to 0.1 µm \cite{Samal2016-js} and reduced interconnect latency. As a result, MIVs can support significantly higher vertical connection densities compared to hybrid bonding or TSVs. However, MIV fabrication requires homogeneous layer stacks, which may limit its applicability in some designs.

Table~\ref{tab:num_conns_and_case_study_1_results} shows the number of vertical connections required for each evaluated architecture. Most architectures exhibit a reasonable number of vertical connections per grid location, remaining below the 302-connection limit achievable with 1 µm pitch hybrid bonding. Specifically, the {3D CB-O}, {3D SB}, and {3D Hybrid-O} architectures require fewer vertical connections, whereas architectures such as {3D CB} and {3D Hybrid} demand higher connection densities due to the large number of inter-layer connections at CLB inputs.
Future work may explore techniques such as routing channel multiplexing to reduce the number of vertical connections.

\para{WL and CPD.}
Figure~\ref{fig:case_study_1_results} and Table~\ref{tab:num_conns_and_case_study_1_results} summarize the WL and CPD results. All 3D architectures outperform the 2D baseline, achieving CPD reductions of 3.1\%–6.7\% and WL reductions of 4.2\%–22.9\%. Among them, the 3D Hybrid architectures deliver the most balanced improvements in both metrics, with the 3D CB architecture achieving the largest WL reduction and the 100\% 3D SB configuration yielding the greatest CPD reduction. While WL improvement shows a clear positive correlation with the number of vertical connections, CPD does not follow the same trend: the 3D SB architectures achieve the best CPD reductions despite having relatively few vertical connections per grid.

\para{Takeaways.} \textbf{(1)} 3D FPGAs show promising improvements in CPD and WL compared to 2D FPGAs. \textbf{(2)} While more vertical interconnects improve WL performance, the relationship with CPD is less straightforward, with diminishing returns at higher connection densities. \textbf{(3)} A balanced architecture that combines different vertical connection types tends to yield the most effective trade-offs.

\input{figures/case_study_1_results}
\input{tables/num_conns_and_case_study_1_results}


\subsubsection{Impact of 3D SB Placement}
\label{sec:case_study_sb_palcement}

\input{figures/sb_placement_results}

This case study investigates the impact of 3D SB placement on CPD, WL, and routing time. Building on the promising results of the 3D SB connection type from Case Study~1, we explore if the spatial distribution of 3D SBs has a significant effect on performance.
We use 3D SB subset connections with 50\% 3D SBs and evaluate six placement strategies, illustrated in Fig.\ref{fig:tool_parameters_e}. The corresponding results are presented in Fig.\ref{fig:sb_placement_results}.
Among the evaluated strategies, the Perimeter placement stands out as a clear outlier, yielding substantially worse CPD and WL than all other approaches and even performing significantly below the 2D baseline. It also incurs an exponentially higher routing time. This degradation is likely due to the VTR placement engine’s lack of awareness of vertical connectivity, which may result in logic blocks requiring vertical links being placed far from the perimeter-located 3D SBs, thereby increasing routing congestion and prolonging runtime. These findings underscore the importance of developing 3D-aware placement algorithms.

Excluding the Perimeter case, the remaining strategies exhibit comparable CPD, WL, and routing times, with the Core placement showing a slightly higher runtime than the others. Strategies with more distributed 3D SBs, such as Repeated Interval, Rows, Columns, and Random, achieve the fastest routing times, while those with concentrated 3D SBs, such as Core and Perimeter, perform worse. This suggests that a more uniform distribution of 3D SBs across the fabric can enable more efficient routing.


\para{Takeaways.} \textbf{(1)} 3D SB placement significantly impacts routing time, even when the number of 3D SBs is held constant. \textbf{(2)} Uniformly distributing 3D SBs across the fabric enables faster and more efficient routing compared to concentrated configurations. \textbf{(3)} The PnR algorithm used in this study may not be optimized for 3D FPGAs and may yield suboptimal results, highlighting the need for 3D-aware PnR strategies.
\input{tables/sb_pattern}
\input{figures/sb_pattern_results}
\subsubsection{Impact of 3D SB Connection Patterns}
\label{sec:case_study_sb_pattern}

This case study investigates the influence of 3D SB connection patterns on performance.  We explored various patterns, outlined in Table \ref{tab:sb_pattern}, including \texttt{revolving} and \texttt{symmetric} patterns inspired by the Wilton pattern commonly used in 2D FPGAs \cite{Wilton1997-bp}. The Wilton pattern, based on an offset with a modulo operation, enhances routability in 2D by reducing the number of turns required to reach desired tracks.  While all vertical connections in 3D inherently involve turns, we investigated whether specific 3D patterns could offer greater benefits compared to others. A randomly generated pattern was also included to show the tool's ability to explore a wide range of patterns.

Results shown in Fig. \ref{fig:sb_pattern_results} indicate that WL and CPD are largely unaffected by the SB pattern. CPD and WL show small variation, with CPD improvements ranging from 4.3\% to 6.7\%, and WL improvements ranging from 10.4\% to 12.1\%. The symettric offset and subset patterns yielded the best CPD results. While further investigation is needed to understand the impact of different patterns, these initial results suggest that the placement of 3D SBs has a greater influence than the connection pattern.

\para{Takeaways.} \textbf{(1)} 3D SB connection patterns appear to have a minimal effect on WL and CPD. \textbf{(2)} The location of 3D SBs appears to be a more significant factor.

\input{figures/vertical_delay_results}
\input{figures/vertical_delay_layer_crossings}

\subsubsection{Impact of Vertical Connection Delay}
\label{sec:case_study_vertical_delay}

This case study examines the effect of vertical connection delay on 3D FPGA performance.  Different 3D integration technologies exhibit varying vertical connection densities and speeds.  To explore this impact, we varied the vertical delay ratio from 0.5$\times$ to 10$\times$ the delay of a length-4 horizontal routing segment. As a reference point, \cite{Boutros2023-og} used a vertical connection delay of 137ps was used (0.739$\times$ the horizontal segment delay), representing hybrid bonding with a 5µm pitch. In our study, a 0.5$\times$ vertical delay corresponds to a vertical delay of 93ps, while a 10$\times$ ratio corresponds to 1.86ns. This wide range encompasses various technologies, from faster hybrid bonding (93ps or less depending on pitch) to slower TSVs (e.g., 1.5ns as reported in \cite{Huang2012-ar}).

Fig. \ref{fig:vertical_delay_results} presents the results.  CPD exhibits a near-linear relationship with the vertical delay ratio.  WL increases slightly for all 3D architectures.  However, a key limitation in VTR's placement engine may influence these results. During initial placement, VTR assigns blocks to layers pseudo-randomly, preferring layers containing connected blocks. Subsequently, the algorithm limits inter-layer movement, with only a $\sim$10\% probability of layer reassignment per block during the placement flow.


This limited layer reassignment potentially limits performance gains, especially considering the interplay between vertical delay and layer crossings. As shown in Fig. \ref{fig:vertical_delay_layer_crossings}, the number of layer crossings remains constant for 3D CB-O architectures and decreases slightly for 3D SB and 3D CB architectures as the vertical delay ratio increases.  This is a direct consequence of VTR's placement limitations. Ideally, with a low vertical delay, designs should maximize layer crossings to exploit vertical connectivity.  Conversely, a high vertical delay should minimize crossings. VTR's limited layer reassignment prevents such dynamic optimization.


\para{Takeaways.} \textbf{(1)} CPD increases approximately linearly with vertical connection delay.  However, VTR's placement limitations may influence this relationship. \textbf{(2)} 3D FPGAs demonstrate performance comparable to their 2D counterparts when the vertical connection delay is approximately two times that of a horizontal connection. This suggests that even with moderately high vertical delays, 3D architectures can remain competitive.

\subsubsection{Impact of Benchmark Placement Algorithm}

To assess the variability of the benchmark PnR process for both 2D and 3D architectures, a study was conducted using randomized placement seeds. For each architecture tested (2D and 100\% 3D SB), 150 unique seeds were generated and used by the placement engine during simulated annealing. The resulting CPD values are presented in Fig. \ref{fig:random_seed_results}.

The results indicate that 3D SB architectures consistently achieve better CPD across all benchmarks. However, the spread of CPD values across different placement seeds is similar for both 2D and 3D architectures. For example, the \texttt{attention\_layer} benchmark shows a wide variation in CPD depending on the seed, whereas the \texttt{conv\_layer} benchmark exhibits a much tighter distribution. This suggests that introducing 3D interconnects does not inherently increase the variation in placement and routing outcomes. Nonetheless, the benchmark-specific differences in CPD variability merit further investigation.


\input{figures/random_seed_results}

%% file: figures/physical_implementation.tex
\begin{figure}[tb]
    \centering
    \includegraphics[width=0.8\columnwidth]{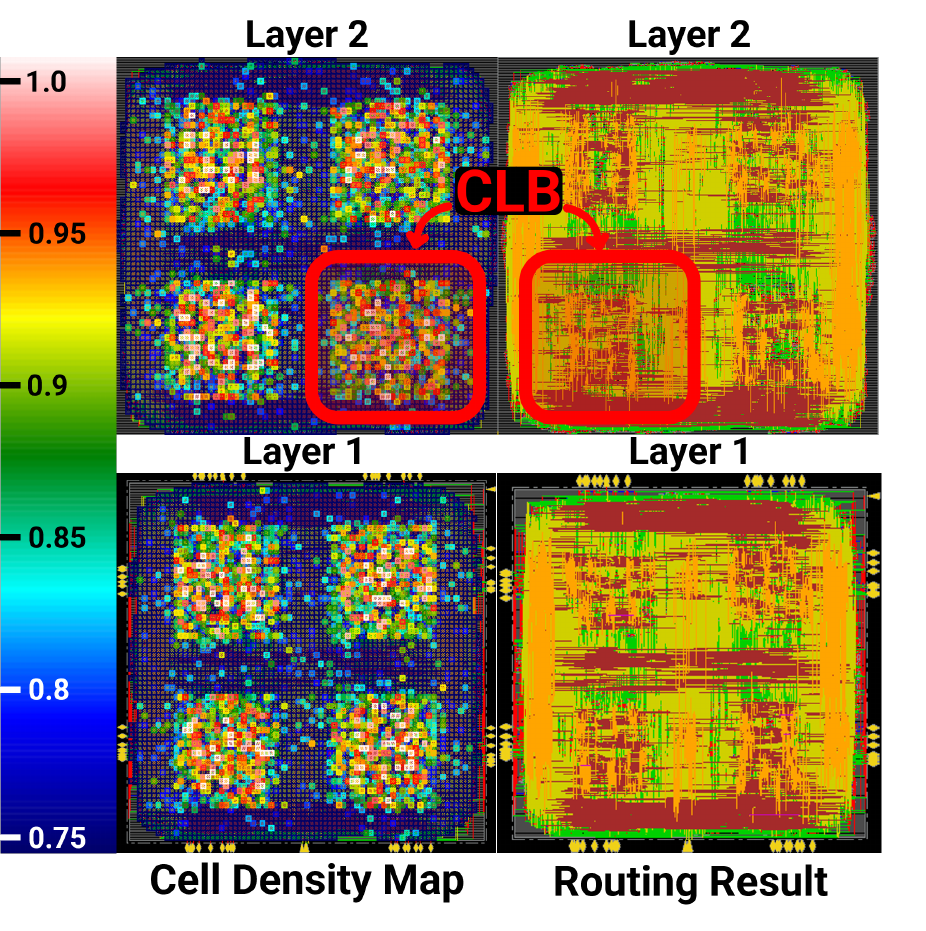}
    \caption{Physical implementation of a two-layer homogeneous 3D-SB FPGA. Routing closure is achieved, verifying \thiswork{}'s RTL generation.}
    \label{fig:pdResult}
    \vspace{4pt}
\end{figure}

%% file: figures/runtime_analysis.tex
\begin{figure}[bt]
    \centering
    \includegraphics[width=0.9\columnwidth]{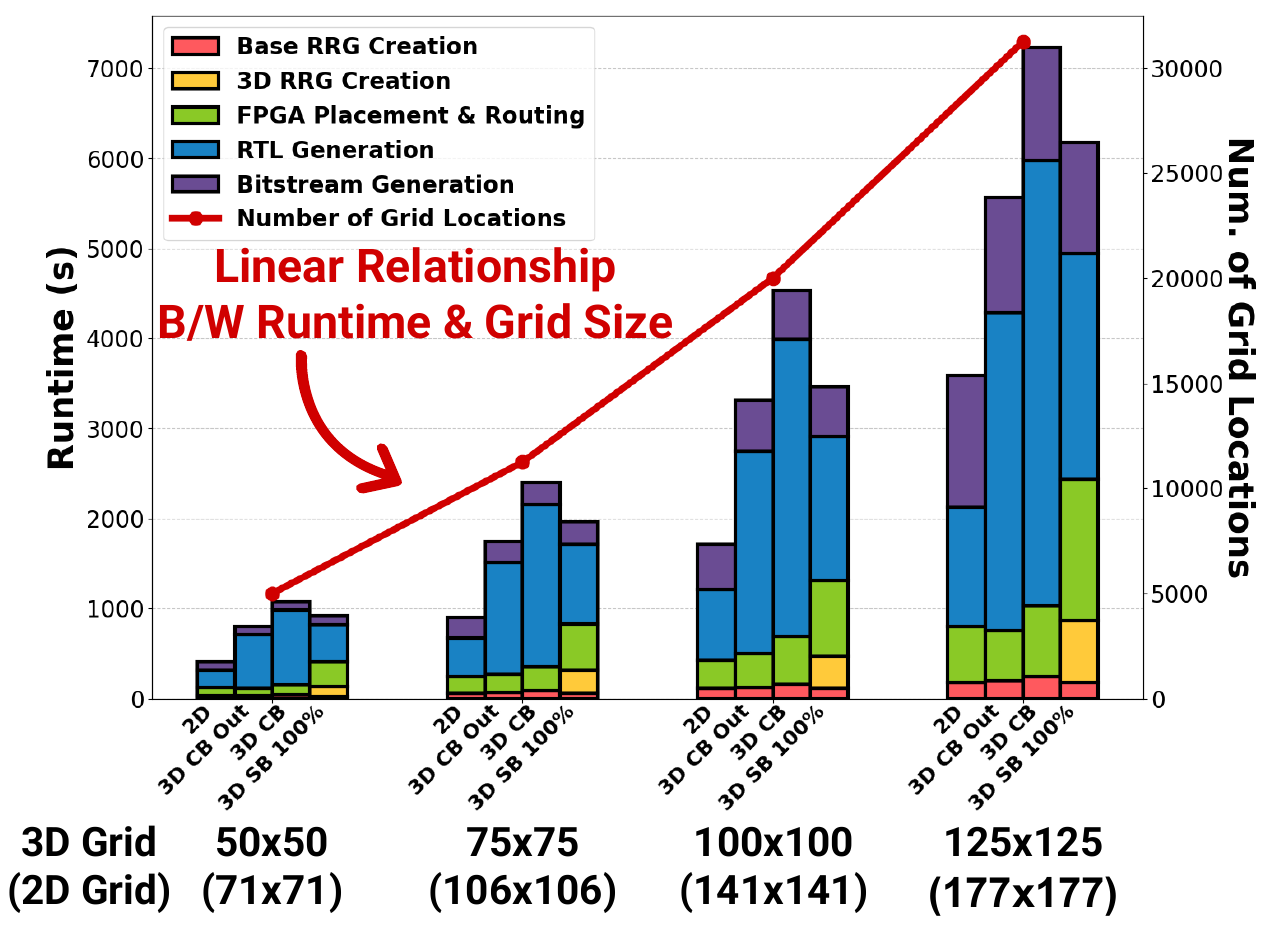}
    \caption{Runtime breakdown of \thiswork{} on various Grid Sizes.} 
    \label{fig:runtime_results}
\end{figure}

%% file: tables/case_studies_parameters.tex
\begin{table}[tb]
    \centering
    \scriptsize
    \setlength\tabcolsep{-1pt}
    \begin{tabular}{C{2.9cm}|C{3.8cm}|C{2.2cm}}
        \toprule
        \rowcolor{gray!25}
        \textbf{Arch. Parameters} & \textbf{Values \& Options} & \textbf{Fixed / Varied} \\
        \hline
        Layer count \& Grid Size & 2 layers; 100\texttimes{}100 (2D: 141\texttimes{}141) & Fixed \\
        Num. of CLBs & 14,308 (2D: 14,317) & Fixed \\
        Num. of DSPs & 2,548 (2D: 2,502) & Fixed \\
        Num. of BRAMs & 2,352 (2D: 2,502) & Fixed \\
        LUT \& Cluster \& CW & LUT: 6; Cluster: 10; CW: 300 & Fixed \\
        Segment Lengths & 4 [260 tracks], 16 [40 tracks] & Fixed \\
        \midrule
        \rowcolor{cyan!15}
        Vertical Connection Types & 3D SB &  Varied in study (1) \\ \hline
        \rowcolor{cyan!15}
        3D SB Percentage & 50\% & Varied in study (1) \\ \hline
        \rowcolor{cyan!15}
        3D SB Placement & Repeated Interval & Varied in study (2)\\ \hline
        \rowcolor{cyan!15}
        SB Patterns & Wilton (horizontal), Subset (vertical) & Varied in study (3) \\ \hline
        \rowcolor{cyan!15}
        Vertical Delay Ratio & 0.739 \cite{Boutros2023-og} (137 ps) & Varied in study (4) \\
        \bottomrule
    \end{tabular}

    \caption{FPGA architectural parameters used in case studies.}
    \label{tab:case_studies_paramters}
\end{table}

%% file: figures/case_study_1_results.tex
\begin{figure*}[t] 
    \centering
    \includegraphics[width=\textwidth]{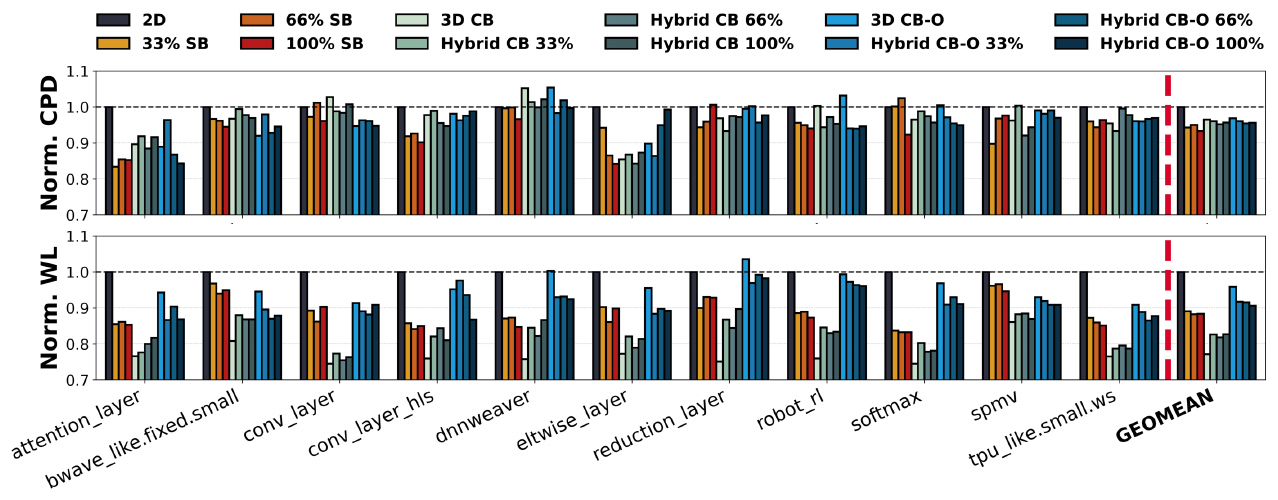}
\caption{Case Study (1): CPD and WL results of various vertical connection types, normalized to 2D FPGAs, using Koios Benchmarks.}
    \label{fig:case_study_1_results}
\end{figure*}

%% file: tables/num_conns_and_case_study_1_results.tex
\begin{table}[tb]
    \centering
    \scriptsize
    
    \setlength\tabcolsep{1.5pt}
    \begin{tabular}{C{2cm}|C{1.3cm}|C{1.5cm}|C{1.6cm}|C{1.6cm}}
        \toprule
        \rowcolor{gray!25} 
        \textbf{Vertical Connection Type} & \textbf{\# of Vert. Connections} & \textbf{Vert. Conns. Per Grid} & \textbf{WL Geomean (\% reduc.)} & \textbf{CPD Geomean (\% reduc.)} \\
        \hline
        3D CB & 3,038,192 & 316.3 & \cellcolor{cyan!15} \textbf{0.771 (22.9\%)} & 0.965 (3.5\%)\\
        3D CB-O & 431,984 & 45.0 & 0.958 (4.2\%) & 0.969 (3.1\%)\\
        3D SB 100\% & 750,680 & 78.2 & 0.884 (11.6\%)& \cellcolor{cyan!15} \textbf{0.933 (6.7\%)}\\
        3D SB 66\% & 495,598 & 51.6 & 0.882 (11.8\%) & 0.950 (5.0\%)\\
        3D SB 33\% & \cellcolor{cyan!15} \textbf{247,832} & \cellcolor{cyan!15} \textbf{25.8} & 0.890 (11.0\%)& 0.943 (5.7\%)\\
        3D Hybrid 100\% & 3,788,872 & 394.5 & 0.827 (17.3\%) & 0.957 (4.3\%)\\
        3D Hybrid 66\% & 3,533,790 & 367.9 & 0.818 (18.2\%) & 0.951 (4.9\%)\\
        3D Hybrid 33\% & 3,286,064 & 342.2 & 0.826 (17.4\%) & 0.960 (4.0\%)\\
        3D Hybrid-O 100\% & 1,182,664 & 123.1 & 0.906 (9.4\%)& 0.956 (4.4\%)\\
        3D Hybrid-O 66\% & 927,582 & 96.6 & 0.916 (8.4\%) & 0.955 (4.5\%)\\
        3D Hybrid-O 33\% & 679,816 & 70.8 & 0.917 (8.3\%) & 0.960 (4.0\%)\\
        
        \bottomrule
    \end{tabular}

     \caption{Impact of vertical connection types. Results normalized to 2D FPGA on Koios benchmark subset.}
     \label{tab:num_conns_and_case_study_1_results}
\end{table}

%% file: figures/sb_placement_results.tex
\begin{figure}[bt]
    \centering
    \includegraphics[width=0.9\columnwidth]{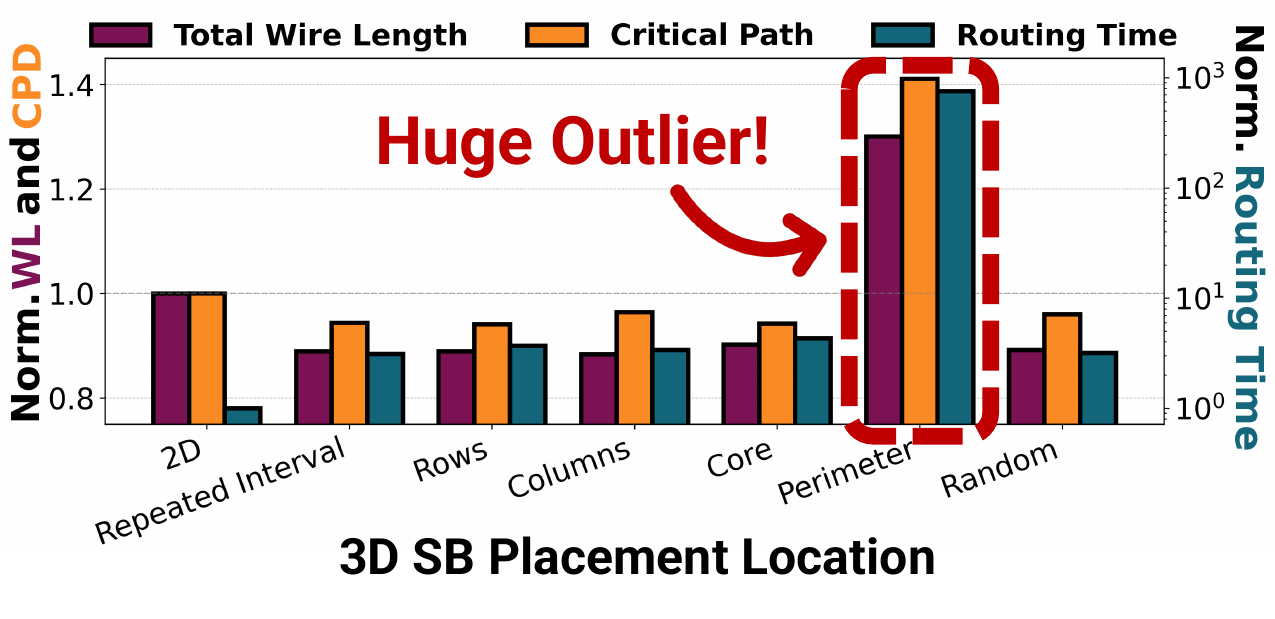}
    \caption{Case Study (2): CPD and WL of different 3D SB Placements.}
    \label{fig:sb_placement_results}
\end{figure}

%% file: tables/sb_pattern.tex
\begin{table}[tb]
    \centering
    \scriptsize
    \begin{tabular}{C{2.5cm}|C{2cm}|C{2cm}}
        \toprule
        \rowcolor{gray!25} 
        \textbf{3D SB Pattern Name} & \textbf{Input Pattern} & \textbf{Output Pattern}\\
        \hline
        Subset & [0,0,0,0] & [0,0,0,0] \\
        Off By One Output& [0,0,0,0] & [1,1,1,1] \\
        Revolving Offset & [0,1,2,3] & [0,1,2,3] \\
        Revolving Input & [0,1,2,3] & [0,0,0,0] \\
        Revolving Output & [0,0,0,0] & [0,1,2,3] \\
        Direction Match & [0,1,0,1] & [1,0,1,0] \\
        Symmetric Offset & [-2,-1,1,2] & [0,0,0,0] \\
        Random & [-3,0,2,1] & [3,-1,-2,2] \\
        \bottomrule
    \end{tabular}
    \caption{3D SB Patterns Tested in Case Study (3).}
    \label{tab:sb_pattern}
\end{table}

%% file: figures/sb_pattern_results.tex
\begin{figure}[bt]
    \centering
    \includegraphics[width=0.9\columnwidth]{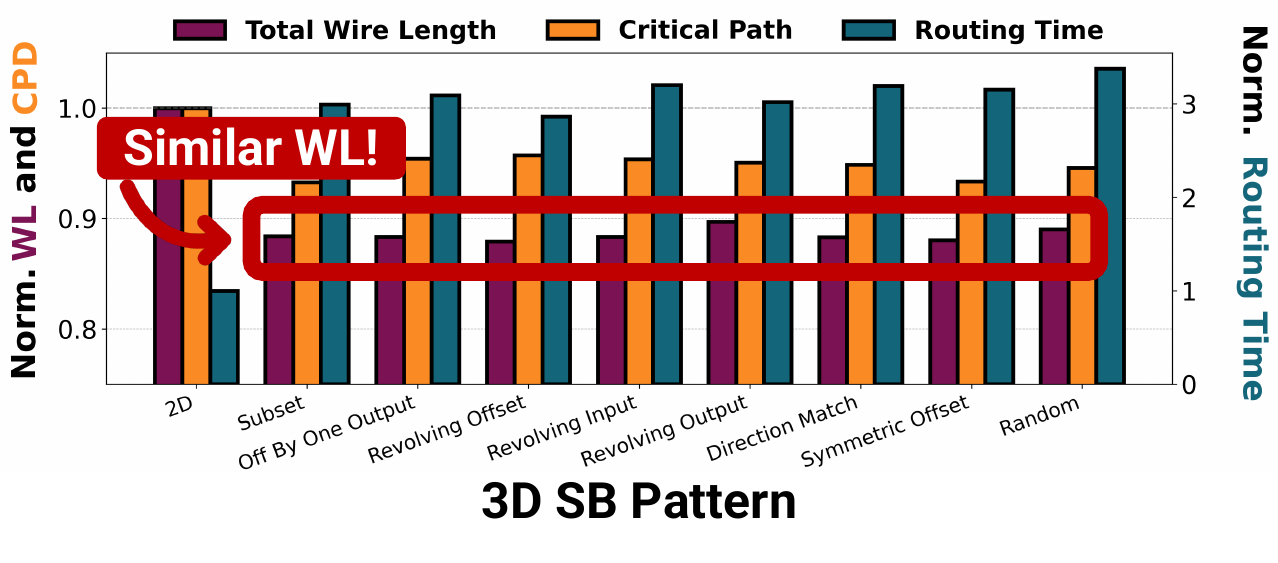}
    \caption{Case Study (3): CPD and WL of different 3D SB patterns.}
    \label{fig:sb_pattern_results}
\end{figure}

%% file: figures/vertical_delay_results.tex
\begin{figure}[bt]
    \centering
    \includegraphics[width=0.8\columnwidth]{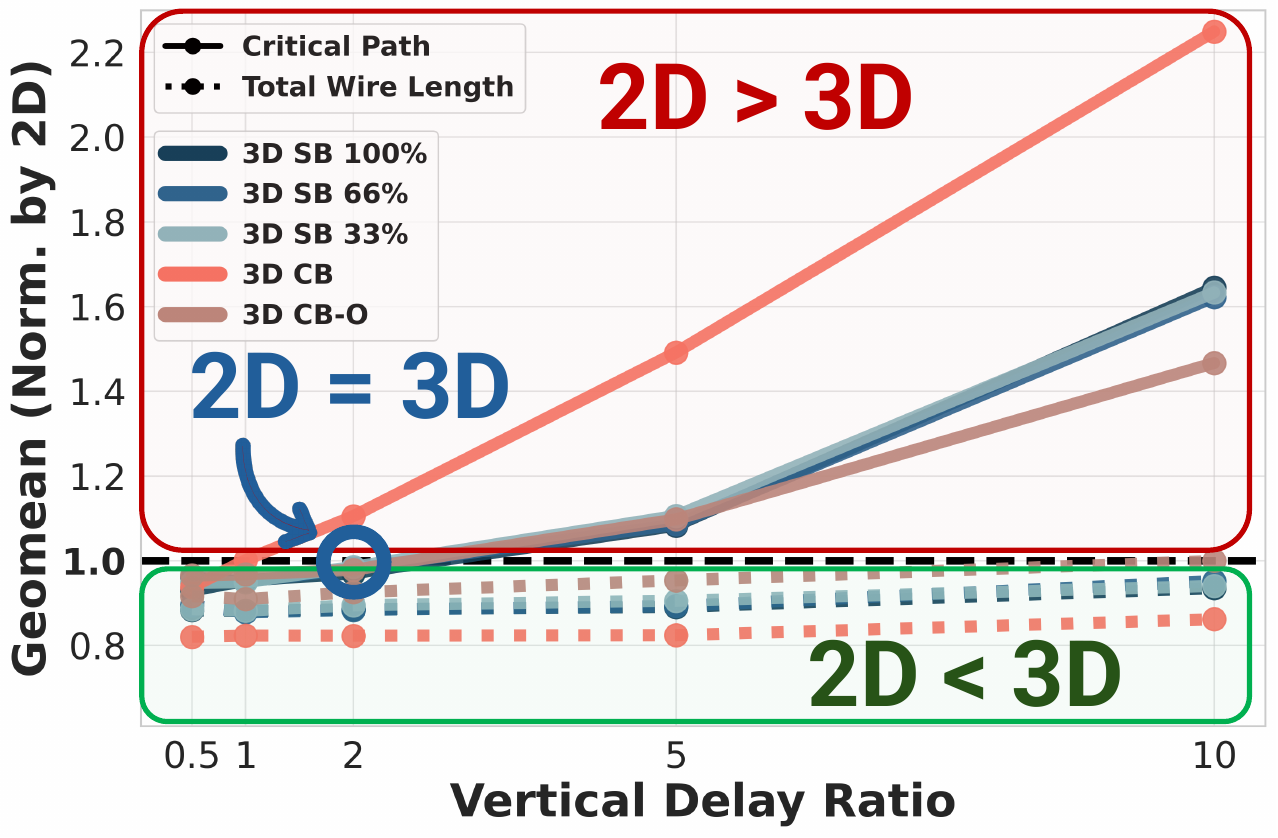}
    \caption{Case Study (4): CPD and WL under different vertical delay ratios, ranging from 0.5 to 10.}
    \label{fig:vertical_delay_results}
\end{figure}

%% file: figures/vertical_delay_layer_crossings.tex
\begin{figure}[bt]
    \centering
    \includegraphics[width=0.9\columnwidth]{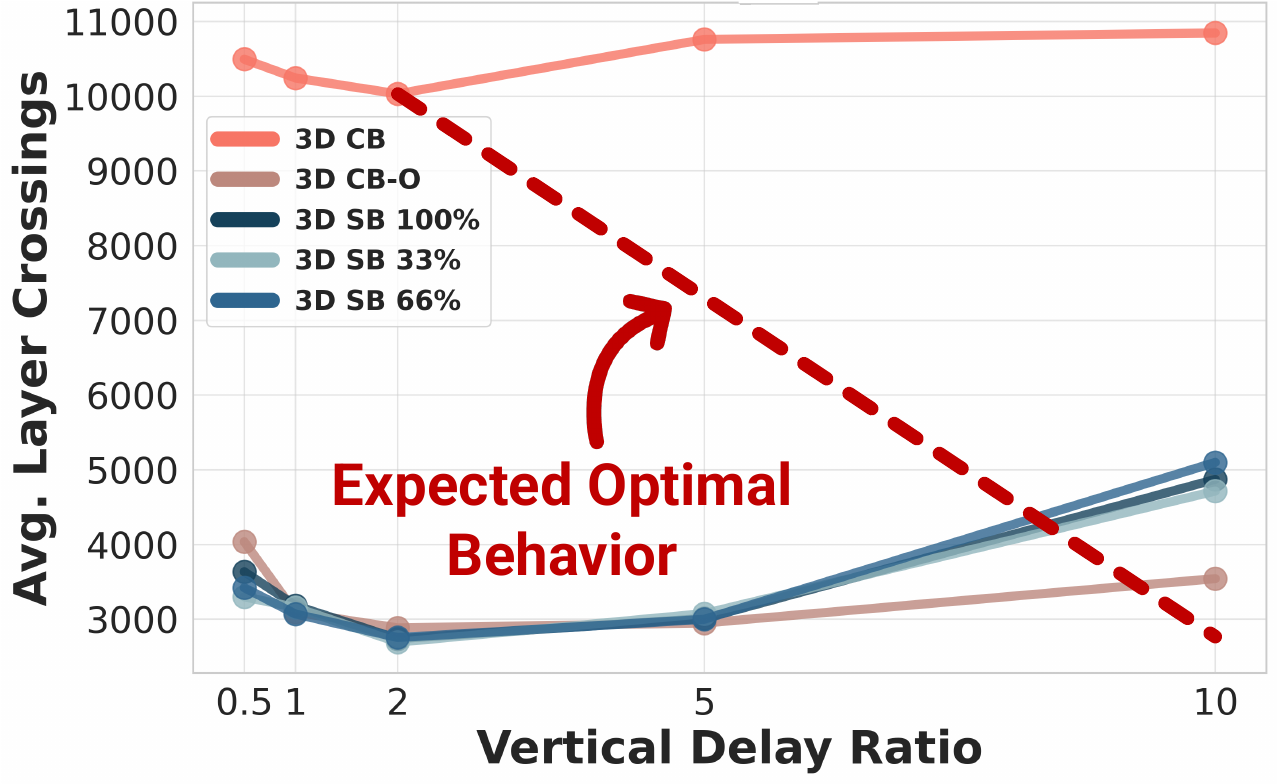}
    \caption{Case Study (4): Average number of signals that cross layers as vertical delay ratio increases.}
    \label{fig:vertical_delay_layer_crossings}
\end{figure}

%% file: figures/random_seed_results.tex
\begin{figure}[tb]
    \centering
    \includegraphics[width=0.9\columnwidth]{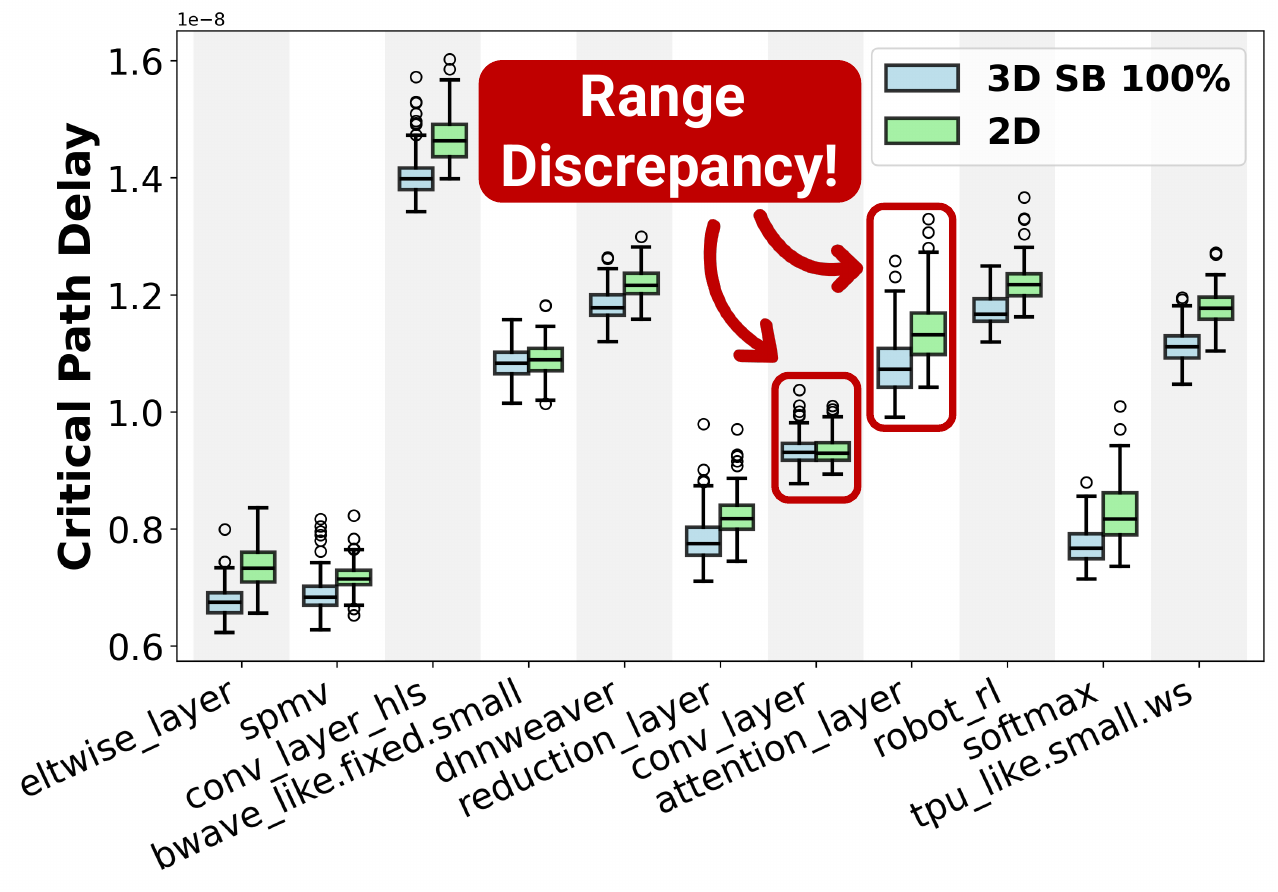}
    \caption{Case Study (5): Distribution of CPD values after 150 random seed placements on 2D and 3D FPGAs with 100\% 3D SBs.}
    \label{fig:random_seed_results}
    \vspace{3pt}
\end{figure}

%% file: sections/5_Conclusion.tex
\section{Conclusion and Future Work}

In this work, we introduced \thiswork{}, an automated tool that streamlines
the design and exploration of 3D FPGA architectures. By generating
custom 3D FPGA fabrics based on user-defined parameters,
 \thiswork{} facilitates efficient exploration of the vast 3D FPGA design
space. And allows the exploration of 3D FPGAs beyond simulation by producing synthesizable RTL that can be physically designed. Our results demonstrate that 3D FPGAs can achieve improvements of up to 6.7\% for critical path delay and up to 22.9\% for total wire length compared to 2D FPGAs.

Looking ahead, we aim to obtain accurate Power, Performance, Area, and Thermal (PPAT) metrics for 3D FPGAs through physical design implementation. Additionally, we plan to investigate placement and routing algorithms specifically tailored for 3D FPGAs, with the potential to further enhance performance.